\documentclass[twocolumn,aps,prl,showpacs,amsmath,amssymb]{revtex4}
\usepackage[dvipdfm]{graphicx}
\usepackage{bm}

\begin{document}

\title{Perfectly Conducting Channel and Universality Crossover 
in Disordered Nano-Graphene Ribbons}
\author{Katsunori Wakabayashi$^1$}
\author{Yositake Takane$^1$}
\author{Manfred Sigrist$^{2,3}$}
\affiliation{$^1$Department of Quantum Matter, AdSM, Hiroshima University,
Higashi-Hiroshima 739-8530, Japan}
\affiliation{$^2$Theoretische Physik, ETH-H\"onggerberg, Z\"urich CH-8093,
Switzerland} 
\affiliation{$^3$ Department of Physics, Kyoto University, Kyoto 606-8502, Japan}

\begin{abstract}
The band structure of graphene ribbons with zigzag edges have two valleys well separated in momentum space, related to the two 
Dirac points of the graphene spectrum. The propagating modes in each valley contain a single
chiral mode originating from a partially flat 
band at band center. This feature gives rise to a perfectly conducting channel in the disordered 
system, if the impurity scattering does not connect the two valleys, i.e. for
long-range impurity potentials. Ribbons with short-range impurity potentials, 
however, through inter-valley scattering 
display ordinary localization behavior. The two regimes belong to 
different universality classes: unitary for long-range impurities and 
orthogonal for short-range impurities.
\end{abstract}

\maketitle

The recent fabrication of graphene devices, combined with observation of half-integer
quantum Hall effect\cite{novoselov} and 
the intrinsic $\pi$-phase shift of the Shubnikov-de Haas 
oscillations\cite{kopelvich}, has once more ignited an intense discussion on
this old fascinating system.  
Due to the two-dimensional honeycomb structure,
the itinerant $\pi$-electrons near
the Fermi energy behave as massless Dirac fermion.
The valence and conduction bands touch conically at two nonequivalent
Dirac points, called $\bm{K_+}$ and $\bm{K_-}$ points,
which possess opposite chirality \cite{chiral}.
\begin{figure}[t]
\includegraphics[width=0.82\linewidth]{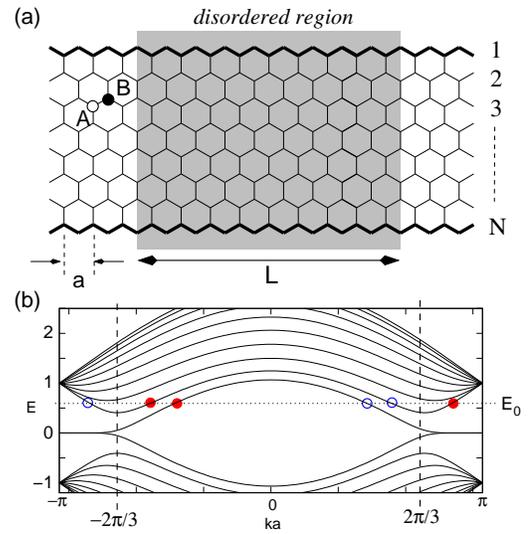}
\caption{(a) Structure of graphene zigzag ribbon. The disordered region with randomly
distributed impurites lies in the shaded region and has
the length $L$. The lattice constant is $a$ and the ribbon width $N$ is defined 
as the number of the zigzag chains.
(b) Energy dispersion of zigzag ribbon with $N=10$. The valleys in the 
energy dispersion near $k=2\pi/3a$ ($k=-2\pi/3a$) originate from the
Dirac $\bm{K_+}$($\bm{K_-}$)-point of graphene.
The red-filled (blue-unfilled) circles denote the right (left)-moving
open channel at the energy $E_0$. In the left(right) valley separately the degeneracy 
between right and left moving channels is missing due to 
one excess right(left)-going mode. The time-reversal symmetry under the
intra-valley scattering is also broken.}
\label{fig:ribbon}
\end{figure}

In graphene, the presence of edges can have strong implications
for the spectrum of the $\pi$-electrons \cite{peculiar}. 
There are two basic shapes of edges, {\it armchair} and {\it zigzag} which
determine the properties of graphene ribbons. 
In ribbons with zigzag edges, localized states appear at the edge with energies close to
the Fermi level \cite{peculiar}. 
In contrast, edge states are absent for ribbons with armchair edges. 
Recent experiments support the evidence of edge localized states\cite{enoki}.
The electronic transport through zigzag ribbons shows a number of 
intriguing phenomena such as zero-conductance Fano resonances \cite{prl},
vacancy configuration dependent transport \cite{vacancy},
valley filtering \cite{rycerz} and half-metallic conduction \cite{son}.

In this Letter, we focus on disorder effects of the electronic transport properties of
graphene zigzag ribbons. The edge states play an important role here, since they
appear as special modes with partially flat bands and lead under certain conditions 
to chiral modes. There is one such mode of opposite orientation 
in each of the two valleys,
which are well separated in $k$-space. The key result of this study is that for disorder
without inter-valley scattering a single perfectly conducting channel emerges
associated with such a chiral mode. This mode disappears as soon as inter-valley scattering
is possible. This distinction depends on the range of the impurity potentials. We will show that
as a function of the impurity potential range a crossover from the orthogonal to the unitary
universality class occurs which is connected with the presence or absence of 
time reversal symmetry (TRS).

We describe the electronic states of
nanographites by the tight-binding model
\begin{eqnarray}
H = \sum_{i,j} \gamma_{i,j}|i\rangle\langle j| 
  + \sum_i V_i |i\rangle\langle i|, 
\label{eq:hamiltonian}
\end{eqnarray}
where $\gamma_{i,j}=-1$ if $i$ and $j$ are nearest neighbors, and 0 otherwise. 
$|i\rangle$ represents the state of the $p_z$-orbital 
on site $i$ neglecting the spin degrees of freedom. In the following we will
also apply magnetic fields perpendicular to the 
graphite plane which are incorporated via the Peierls phase: 
 $\gamma_{i,j}\rightarrow\gamma_{i,j}\exp\left[
i2\pi(e/ch)\int_i^jd\bm{l\cdot A}\right]$, where $\bm{A}$
is the vector potential. 
The second term in Eq. (\ref{eq:hamiltonian}) represents the 
impurity potential, 
$V_i=V(\bm{r}_i)$ is the 
impurity potential at a position $\bm{r}_i$.

\begin{figure}[h]
\includegraphics[width=0.7\linewidth]{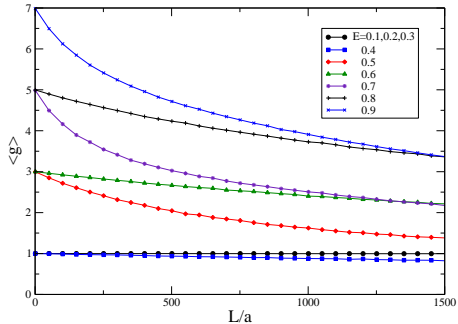}
\caption{$L$-dependence of the
average of dimensionless conductance $ \langle g \rangle $ for zigzag ribbon
with $N=10$,  $d/a=1.5$ (no inter-valley scattering),  $u_0=1.0$, and
$n_{imp.}=0.1$. 6000 samples with different impurity configuration
are included in the ensemble average.}
\label{fig:aveg}
\end{figure}

As shown in Fig.\ref{fig:ribbon}(a), our zigzag ribbons are characterized 
by the width $N$, the number of  zigzag chains, and $L$ denotes 
the length of the disordered region. In Fig. \ref{fig:ribbon}(b), we display the
band structure for the zigzag ribbon with $N=10$.  Note that zigzag 
ribbons are metallic for all widths at finite doping because of the presence of
a partial flat band at zero energy induced by edge states.
These edge states lead in the clean
limit to the characteristic conductance odd-number quantization, i.e. $g=2n+1$ as the dimensionless conductance per spin ($ n =0,\pm1,\pm2, \dots$) \cite{prl,peres}.
There are two {\it valleys}, at
$ k_{\pm} = \pm 2 \pi /3 $, each of which possesses one excess mode which violates the balance between the number left-
and right-moving modes (Fig.\ref{fig:ribbon}).

\begin{figure}
\includegraphics[width=\linewidth]{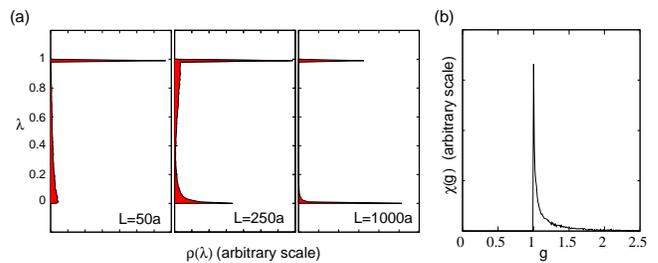}
\caption{(a) Distribution of the transmission eigenvalues $\lambda$: $\rho(\lambda)$,
at $E=0.5$ for $L/a=50,250,1000$, with $d/a=2.0$. $E=0.5 $ leads to 3 incident channels.
12,000 samples with different impurity configurations are included in the distribution.
(b) Distribution of the dimensionless conductance $g$: $\chi(\lambda)$, at $L/a=1000$ 
 for the same parameter set. } 
\label{fig:eigen5}
\end{figure}

In our model  we assume that the impurities are randomly distributed with a density $n_{imp}$, and
the potential has a Gaussian form of range $d$
\begin{equation}
V(\bm{r}_i) = \sum_{\bm{r_0}(random)}u
\exp\left(-\frac{|\bm{r}_i-\bm{r}_0|^2}{d^2}\right) 
\end{equation}
where the strength $u$ is uniformly distributed within the range $|u|\le u_{M}$.
Here $u_{M}$ satisfies the normalization condition:
$u_{M}\sum_{{\bm r}_i}^{(full\ space)}
\exp\left(-{\bm{r}_i^2}/{d^2}\right)/(\sqrt{3}/2)=u_0$.
The range of the impurity potential is crucial for the transport properties.
Since the momentum difference between two
valleys is rather large, $\Delta k = k_+ - k_- =4\pi/3a$, 
only short-range impurities (SRI) with a range smaller than
the lattice constant causes {\it inter-valley scattering}. Long-range impurties (LRI), in contrast,
restrict the scattering processes to {\it intra-valley scattering} \cite{ando}.

We briefly discuss here the relation between valleys in the zigzag ribbons and graphene.
The electronic states near the Dirac point can be
described by the $\bm{k\cdot p}$ Hamiltonian 
\begin{equation}
H_{\bm{k\cdot p}}= \tilde{\gamma}
\left[
\hat{k}_x(\sigma^x\otimes \tau^0 )
+\hat{k}_y(\sigma^y\otimes\tau^z)
\right]
\end{equation}
acting on the 4-component pseudo-spinor Bloch functions
$\Phi=\left[
\phi_{\bm{K_+}A},
\phi_{\bm{K_+}B},
\phi_{\bm{K_-}A},
\phi_{\bm{K_-}B}
\right]$, which characterize the wave functions on the two crystalline
sublattices (A and B) for the two Dirac points (valleys) $ \bm{K}_{\pm} $.
Here, $\tilde{\gamma}$ is the band parameter, $\hat{k}_x$($\hat{k}_y$)
are wavenumber operators, and $\tau^0 $ is the $2 \times 2 $ identity matrix.
Pauli matrices $\sigma^{x,y,z}$ act on the sublattice space ($A$, $B$),
while $\tau^{x,y,z}$ on the valley space ($\bm{K}_{\pm}$).
Since the outermost sites along $1^{st}$ ($N^{th}$) zigzag chain
are B(A)-sublattice, an imbalance between two sublattices occurs
at the zigzag edges leading to 
the boundary conditions
\begin{equation}
\phi_{\bm{K_\pm}A}(\bm{r}_{[0]})=0, \quad \phi_{\bm{K_\pm}B}(\bm{r}_{[N+1]})=0,
\end{equation}
where $\bm{r}_{[i]}$ stands for
the coordinate at $i^{th}$ zigzag chain.
It can be shown that
the valley near $k=3\pi/2a$ in Fig.1(b) originates
from the $\bm{K_+}$-point, the other valley at $k=-3\pi/2a$
from $\bm{K_-}$-point \cite{Phd}. 
\begin{figure}
\includegraphics[width=0.7\linewidth]{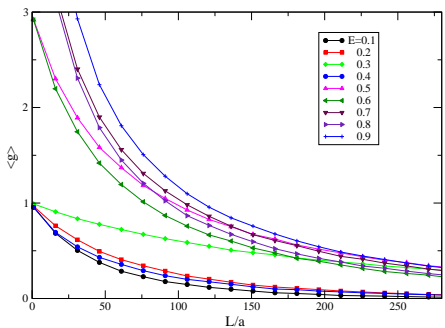}
\caption{$L$-dependence of the
averaged dimensionless conductance for zigzag ribbons with
$N=10$, short-ranged impurity potentional ($d/a=0.05$, inter-valley scattering), 
$u_0=1.0$, and $n_{imp.}=0.1$. 5000 samples with different impurity configurations
are in the ensemble average.}
\label{fig:aveg_short}
\end{figure}

Ignoring the spins the graphene systems have TRS with respect to
the operator $\mathcal{T}$ represented by the complex conjugation $\mathcal{C}$.
Pairs of time reversed states are formed across the two-valleys (Dirac points) as expressed in the above pseudo-spinor formulation by $\mathcal{T}=-i(\sigma^y\otimes\tau^0)\mathcal{C}$, where the A-B-sublattices act as pseudospin degrees of freedom. 
The boundary conditions which treat the two sublattices asymmetrically leading to edge states
give rise to a single special mode in each valley. Considering now one of the two 
valleys separately, say the one around  $ k=k_+ $ we see that the TRS is effectively violated in the sense that we find one more left-moving than right-moving mode.
 If we restrict ourselves
to disorder promoting only intra-valley scattering, transport properties would resemble
those of a system with a chiral mode which is oppositely oriented for the two valleys. In this sense
such a system violates TRS. On the other hand, disorder yielding inter-valley
scattering would restore this TRS as both valleys
together incorporate a complete set of pairs of time-reversed modes. 
Thus we expect to see qualitative differences in the properties if the range of the impurity potentials
is changed.

In order to demonstrate this we now turn to the discussion of the transport properties. 
The dimensionless electrical conductance is calculated using 
the Landauer-B\"uttiker formula\cite{mclbf},
$g(E) = {\rm Tr}(\bm{t}\bm{t}^\dagger),$
where $\bm{t}(E)$ is the transmission matrix
through the disordered region. This transmission matrix can be calculated by means of
the recursive Green function method \cite{green,prl}. We focus first on the case of LRI
using a potential with $ d/a=1.5 $ which is already sufficient to avoid inter-valley scattering. 
Fig.\ref{fig:aveg} shows the averaged dimensionless conductance as a function of $ L $ for different incident energies, averaging over an ensemble of 6000 samples with
different impurity configurations for ribbons of the width $N=10$.
The potential strength and impurity density are chosen to be 
$u_0=1.0$ and $n_{imp.} = 0.1$, respectively. As a typical localization effect we observe
that $\langle g \rangle $ gradually decreases with growing length $ L $ (Fig.\ref{fig:aveg}).
Surprisingly, the ribbons remain highly conductive even at the 
length of $L=1500a$, i.e. more than $350nm$ in the real system. Actually, $ \langle g \rangle $ 
converges to $\langle g\rangle =1$, indicating the presence of a single {\it perfectly conducting} 
channel. It can be seen that $\langle g\rangle(L) $ has an exponential behavior as
$\langle g\rangle -1 \sim \exp(-L/\xi)$ with $\xi$ as the localization length\cite{defxi}. 

We performed a number of tests to confirm the presence of this perfectly conducting channel. 
First of all, it exists up to $L=3000a$ for various ribbon widths up to $N=40$ for the
potential range ($d/a=1.5$). Moreover it remains for LRI with $d/a=2.0,4.0,6.0,8.0$, and $u_0=1.0$,  $n_{imp.}=0.1$ and $ N=10 $. As the effect is connected with the subtle feature of an excess mode in band 
structure, it is natural that the result
can only be valid for sufficiently weak potentials. For potential strengths comparable to the
energy scale of the band structure, e.g. the energy difference between the transverse modes, 
the result should be qualitatively altered\cite{vacancy}. Deviations from the limit $ \langle g \rangle \to 1 $ 
also occur, if the incident energy lies at a value close to the change between $ g = 2n-1 $ and $ g=2n+1 $ 
for the ribbon without disorder. This is for example visible in above calculations for $E =0.4 $
where the limiting value $ \langle g \rangle < 1 $ (Fig.\ref{fig:aveg}). 
As a further test we evaluate the distribution of the transmission eigenvalues
and dimensionless conductance for fixed wire length.
In Fig.\ref{fig:eigen5}(a), the distribution of the eigenvalues $\lambda$ of
the Hermite matrix, $\bm{tt}^\dagger$, is depicted for 
various wire lengths. With growing length $ L $ a progressive separation of the transmission 
eigenvalues emerges with a strong peak close to 0 (localization) and at 1 (perfect conduction channel).
The distribution of the conductance $ g $ (trace of the transmission matrix  
${\rm Tr}(\bm{tt}^\dagger)$), is depicted in Fig.\ref{fig:eigen5}(b)  for samples in the long-wire limit.
Obviously, $ g $ only distributes above $g=1$ with a singularity at 1. 
\begin{figure}
\includegraphics[width=\linewidth]{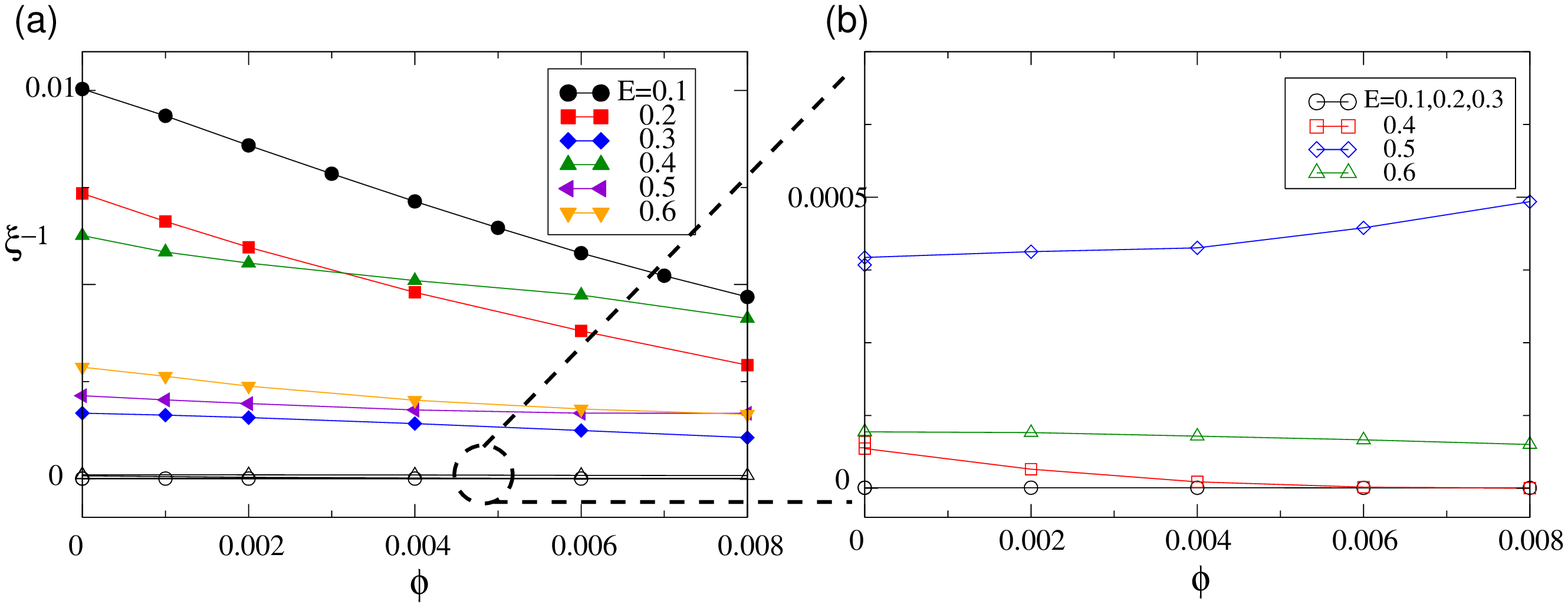}
\caption{(a) Magnetic field dependence of the inverse
 localization length $\xi^{-1}$ for various incident energies.
The filled symbols indicate data sets of systems with SRI being rather sensitive to
magnetic fields, and empty symbols denote data sets for LRI which are almost
insensitive to the field. The magnetic flux $ \phi$ passing through a single
hexagon ring is measured in units of the quantum flux ($\phi_0=ch/e$).
(b) Enlarged view of (a) of LRI data 
5000 samples with different configurations are included in the averages.}
\label{fig:xi}
\end{figure}

Turning to the case of SRI the inter-valley scattering becomes sizable enough
to ensure TRS, such that the perfect transport supported by the
effective chiral mode in a single valley ceases to exist. SRI causes true back scattering. 
For a comparison, we show the ribbon length dependence of the averaged conductance in 
Fig.\ref{fig:aveg_short}. For any incident energy the electrons tend to be localized and 
the averaged conductance decays exponentially, $\langle g\rangle\sim \exp(-\xi/L)$, without
developing a perfect conduction channel.

In order to demonstrate that the qualitative difference between the two regimes, LRI and SRI, is
indeed connected with TRS, we study the effect of magnetic field coupling to
the electrons through the Peierls phase. For the time reversal symmetric situation resulting from SRI
scattering the magnetic field removing TRS should have a stronger effect than for the
case of LRI where TRS is broken already at the outset. 
We use the localization length $ \xi $ as an indicator. In Fig.\ref{fig:xi}, the field 
dependence of the inverse localization length is shown for various incident energies
(filled symbols for SRI and empty symbols for LRI). Indeed the localization length displays a stronger
field dependence than the LRI. Actually for LRI even a so-called anti-localization behavior with increasing field is visible consistent 
with recent reports on graphene\cite{antilocalization,suzuura.prl,wu}.
Note that for $ E < 0.4 $ only a single channel is involved in the 
conductance such that for LRI no localization occurs, i.e. $ \xi^{-1} = 0 $. 

The presence of one perfectly conducting channel 
in disordered quantum wires with symplectic symmetry and an
odd number of channels has recently been analyzed
using random matrix theory \cite{takanesakai}. 
The symplectic symmetry of such systems is based on the 
skew-symmetry of the reflection matrix, $^t\bm{r}= -{\bm r}$  \cite{suzuura}. A
realization can be found in the disordered metallic carbon nanotubes 
with LRI. On the other hand, zigzag ribbons without inter-valley scattering are
not in the symplectic class, since they break TRS in the
special way. The decisive feature for a perfectly conducting channel is
the presence of one excess mode in each valley.
Note that this is 
in contrast to graphene for which each mode has a partner mode of reversed velocity 
in the same valley. 
For single valley transport the reflection 
matrix has a non-square form ($N^{(r)}_c \times N^{(t)}_c$ with $N^{(r)}_c=N^{(t)}_c+1$, 
where $N^{(r)}_c(N^{(t)}_c)$ is the number of
the reflection (incident) channels). 
Recently Hirose {\it et. al.} pointed out that non-square 
reflection matrices with unitary symmetry give rise to a perfectly
conducting channel \cite{ohtsuki}.

Eventuelly we can identify the universality classes of zigzag ribbons. 
For LRI they belong to the {\it unitary} class 
(no TRS), while for
SRI with inter-valley scattering they are in the {\it orthogonal} class 
(with overall TRS). This classification is 
compatible with the behavior in a magnetic field. 

Analogous symmetry considerations can be applied to armchair ribbons. In this case
the two valleys merge into a single one at $k=0$. TRS is conserved
irrespective of the impurity potential range, if there is no magnetic field.
Consequently, disordered armchair ribbons belong always to the
orthogonal class and do not provide a perfectly conducting channel.
In view of the fact that graphene is known to be symplectic (orthogonal) for LRI (SRI) \cite{suzuura.prl}, it is
quite intriguing to realize that the edges influence the universality
class, as long as the phase coherence length is larger than the system
size of nanographenes. 

The unusual energy dispersion due to their edge states gives rise to 
the unique property of zigzag ribbons. Concerning transport properties for disordered
systems the most important consequence is the presence of a perfectly conducting channel.
The origin of this effect lies in the single-valley transport which is dominated by a chiral mode. On the other hand, large momentum transfer through impurities with short-range potentials involves both valleys, destroying this effect and leading
to usual Anderson localization.  The obvious relation of the chiral mode with 
time reversal symmetry leads to the classification into the unitary and orthogonal class
depending on the range of impurity potential. Since the 
inter-valley scattering is weak in the experiments of graphene, we may assume that these
conditions may be realized also for ribbons. Naturally defects  in the 
ribbon edges and vacancies would be
rather harmful for the experiment making this type of experiment very challenging \cite{prl}.

We thank T. Enoki, K. Kusakabe and T. Ohtsuki for stimulating discussions.
This work was financially 
supported by the Swiss Nationalfonds through Centre for Theoretical Studies of
ETH Zurich and the NCCR MaNEP, also supported by a Grand-in-Aid for
Scientific Research (C)  
from the Japan Society for the Promotion of Science (No. 16540291).
The numerical calculation was performed on the Grid/Cluster
Computing System at  Hiroshima University.


\begin{references}
\bibitem{defxi}
In this paper, the localization length is evaluated as
$\xi^{-1}=- d\langle ln \tilde{g}\rangle/dL$. 
Here, $\tilde{g}=g-1$ ($\tilde{g}=g$)
for the system with (without) the perfectly conducting channel.
\end{references}
\end{document}